\documentclass{llncs}
\usepackage{multicol}
\usepackage{pslatex}
\usepackage{amsmath}
\usepackage{amssymb}
\usepackage{graphicx}
\usepackage{mathrsfs}
\usepackage{algorithmic}
\usepackage{algorithm}
\usepackage{subfig}
\usepackage{url}
\usepackage[left=1in,top=1.2in,right=1in,bottom=1in]{geometry}

\title{TrackMeNot-so-good-after-all}
\author{Rami Al-Rfou'
\and William Jannen
\and Nikhil Patwardhan \\
\{ralrfou, wjannen, npatwardhan\}@cs.stonybrook.edu}
\institute{Department of Computer Science \\
  Stony Brook University \\
  NY 11794, USA
}
\begin{document}

\maketitle

\pagestyle{headings}

\begin{abstract}
  TrackMeNot is a Firefox plugin with laudable intentions - protecting
  your privacy. By issuing a customizable stream of random search
  queries on its users' behalf, TrackMeNot surmises that enough
  ``search noise'' will prevent its users' true query profiles from
  being discerned. However, we find that clustering queries by
  semantic relatedness allows us to disentangle a nontrivial subset of
  true user queries from TrackMeNot issued noise.
\end{abstract}

\begin{multicols}{2}
\section{Background}
\label{sec:background}

The decreasing cost of persistent media for storage, coupled with the
steady rise in E-commerce, social networking, and various other online
services, has led to a dramatic increase in the volume of readily
available, personally identifiable information. One often overlooked
source of such information is the logging of user queries by search
engines.

There have been several high-profile examples of search query data
being used in inappropriate ways, most notably an incident involving
AOL in 2006 \cite{aol}. The company disclosed a data set comprised of
information from 658,000 subscribers' search histories; it was
released as a flat file into the public domain. Upon admitting their
error, they removed the link, but the data was already mirrored
elsewhere. This incident highlights just how personally revealing
search engine queries can be, as at least one user was identified
using only the content of her searches \cite{user4417749}.

The average Internet user may browse under the false impression that
deleting cookies or diverting traffic through a proxy will protect
their anonymity on the web. Neither of these measures provide complete
protection. If a user wishes to utilize certain services such as
Google Mail, they must log in to their personal profile. Once logged
in, their queries are then tied to their account.

TrackMeNot is a Firefox plugin that aims to protect the privacy of its
users by issuing random queries on their behalf. These random queries
are pulled from a variety of sources, with the intuition that
providing enough ``noise'' around a user's true search patterns will
make it impossible to disentangle the queries made by TrackMeNot from
those actually made by human searchers.

\section{TrackMeNot details}
\label{sec:tmn}
TrackMeNot operates completely on the client side as a Firefox
plugin. Upon start-up, an initial {\it seed list} is populated from
RSS feeds and known public ``popular search'' lists. During execution,
queries are selected from this list and issued to designated search engines
on the user's behalf.

The user is offered many parameters, which can be tuned to customize
TrackMeNot behavior. For instance, the user can select which, if any,
popular search engines he wishes TrackMeNot to query on his behalf. If
the user only performs searches at \url{http://www.Google.com}, he can
specify that TrackMeNot query only that site as well. The user can
also specify the frequency of TrackMeNot queries (the default average
query rate is 10 queries per hour), as well as whether or not to
enable {\it query bursts}. Query bursts are triggered by actual user
queries; when TrackMeNot detects a genuine query, it responds by
issuing a batch of simulated searches. In a subset of query bursts, a
longer query is selected and permuted to form a set of potentially
similar searches.

The basis of the TrackMeNot model is its {\it dynamic query
  list}. Starting from the initial seed list, queries are randomly
marked for replacement over time. When a marked query is sent by TrackMeNot as a
search engine request, the HTTP responses are parsed to identify any
suitable ``query-like'' terms for replacement. If an acceptable term
is found, it is then substituted into the dynamic list in exchange for
the marked query. In this way, the list of searches will evolve over
time.

TrackMeNot also attempts to simulate user browsing patterns with
``selective click-through''; upon issuing a random query,
it will parse the results page, using regular expression matching to
identify and remove revenue generating ads. It then selects one or
more of the remaining links on the page and simulates a user click.

Although TrackMeNot simulates temporal search patterns with techniques
like query bursts and selective click-through, it may not simulate
contextual search patterns - sequential TrackMeNot queries are most
likely semantically unrelated. Individual queries are selected at
random from a 100-200 term dynamic query list. So while it is likely
that the dynamic query list contains related searches, as long as
searches are selected individually at random, there is no guarantee
that related terms will be issued in temporal proximity. This is the
aspect of the TrackMeNot design we wish to exploit.

\section{Design overview}
\label{sec:design}
Our overall design can be logically divided into three parts. The
first part deals with logging all user Google Search queries as well
as those fired from TrackMeNot. The second part comprises the
computation of a similarity matrix on this set of queries using a
suitable semantic distance measure. The last part consists of
clustering this data using a suitable technique to get a visual
representation of the data from which conclusions can be deduced.

\section{Data Gathering}
To gather the data we ran TrackMeNot with the default behaviors of 10
queries per hour with burst queries enabled; we also enabled the
logging feature and disabled the other search engines, focusing on
Google only. Moreover, we disabled the instant search feature as it
produces incomplete queries. However, this does not give Google a
chance to distinguish TrackMeNot queries as they appear as inserted by
the keyboard. On the Google side, we made a new Google account,
\textbf{cse509tmn}, and enabled the web history for verification.

To collect and label queries, we implemented a proxy server to log all
HTTP requests made by the browser. By collating the proxy log
with the TrackMeNot log, we were able to differentiate between the
user queries and those generated by TrackMeNot.

The plugin was running for at least 3 days on three machines with
three different users. Data was compiled over different periods for
each user.

\section{Semantic Distance Measure}
The basic idea here was to find how \textit{similar} each search query
was to every other query, making no distinction based on its origin
(user or TrackMeNot). To do this, we needed a \textit{semantic}
distance measure. We did not implement our own measure, but instead
explored the available choices from other research groups. After
examining the feasibility of using different available measures, we
settled on two of them. One of them is
DISCO\footnote{http://www.linguatools.de/disco/disco\_en.html}, which
is a Java package, and the other is Google Normalized Distance. For
details, see sections \ref{sec:disco} and \ref{sec:gnd}
respectively. In each of these cases, we disabled the auto-complete
feature of Google so that our search queries were always the exact
phrases that the user or TrackMeNot requested from the Google Search
Engine and never prefixes of the actual queries.

\subsection{DISCO}
\label{sec:disco}
The DISCO (extracting DIStributionally related words using
CO-occurrences) API allows to compute the semantic similarity between
any two words by looking up those words in a pre-defined
repository. For our analysis, we downloaded the Wikipedia repository
available on the DISCO website to compute the distances. By looking up
this local repository, the DISCO API returns a value between 0 and 1
for any pair of words that it looks up in the repository, such that
the higher the value returned, the more similar the words are. In this
sense, the API works like a similarity measure. We were however faced
with one issue: Google search queries are typically phrases, and not
single words. To overcome this, given two queries $Q1$ and $Q2$, we
compared each word in $Q1$ with every word in $Q2$ and in each case
chose the highest returned value. We aggregated this score over all
words in $Q1$ and normalized the addition by dividing the aggregate
score by the number of words in $Q1$.

\subsection{Google Normalized Distance}
\label{sec:gnd}
Google Normalized Distance (GND) is a measure of semantic similarity
based on statistics. GND utilizes an extremely large vocabulary
set. The similarity can be calculated according to the following
equation:

\begin{eqnarray}
\label{eq:gnd}
GND(Q1, Q2)=
&\frac{Max(\log f(Q1), \log f(Q2))-\log f(Q1, Q2))}{\log N - Min(\log f(Q1), \log f(Q2))}
\end{eqnarray}

where $f(Q)$ is the number of results found by Google for query $Q$,
and $f(Q1, Q2)$ denotes the number of results found by Google for the
combined query of $Q1$ and $Q2$.  The more often two queries appear
together, the smaller the distance between them. GND is not an
accurate function to measure the distance\cite{DBLP:GND}. Moreover,
GND is not a metric nor does it satisfy the triangle inequality, and
$GND(x, y)$ is not larger than 0 for every $x \neq y$.

\section{Disentangling user queries}
\label{sec:disentangle}
Once we create a matrix of semantic similarities, we are tasked
with deciding which queries were actually made by the user. We first
perform cluster analysis to identify groups of related queries. We
then use the size and distribution of the cluster assignments to classify
a set of user queries.

\subsection{Clustering}
\label{sec:clustering}
For clustering we use the Partitioning Around Medoids algorithm
(PAM). It is conceptually similar to the well known $k$-means
algorithm; however, rather than minimize squared Euclidean distances,
it aims to minimize dissimilarity \cite{Kaufmann1990}. 

An ideal assignment of $n$ objects into clusters would be an
assignment that minimizes cluster widths while maximizing cluster
separations. In other words, each element $i$ belonging to a cluster
$A$ should be closely related the other members of cluster $A$. At the
same time, each element $i$ in cluster $A$ should be unrelated to the
members of all other clusters $C$ such that $C \neq A$; individual clusters
should be well separated from other clusters.

Formally, for an element $i$ in a cluster $A$, let $a(i)$ denote the
average dissimilarity between $i$ and all elements $j \in A$ where $j\neq
i$. Let $d(i,C)$ denote the average similarity between an element $i$
and all element's in a cluster $C$ to which $i$ does not belong: $C
\neq A$. If we let $b(i)$ be the minimum $d(i,C)$ such that $C \neq
A$, then

\[ s(i) = \frac{b(i)-a(i)}{\max{(a(i),b(i))}} \]

\noindent is a good indicator to the quality of element $i$'s cluster
assignment. If $s(i)$ is close to one, then $i$ is well assigned. If
$s(i)$ is close to negative one, then $i$ is poorly assigned. $s(i)$
is known as the Silhouette width. Choosing the number of clusters, $k$,
that maximizes the average Silhouette width is the heuristic used to
determine the cluster assignments \cite{Kaufmann1990}.

For our implementation of the PAM algorithm, we use the cluster
package from the R statistical programming language\cite{Rcluster}.

\subsection{Classification}
\label{sec:classification}
Our classification strategy is a simple one. It is not our goal to
show that every user query can be discerned from those queries issued
by TrackMeNot. What we want to show is that some nonempty subset of
user queries can be identified as ``real'' with high probability. For
this task, we simply choose the single largest cluster (or set of
clusters in the case of a tie), and label all constituent members as
user-generated queries.

\subsection{Validation}
We have compiled and performed testing on four data sets, containing
both user and TrackMeNot generated searches. The data sets are
summarized in the following table.

\bigskip
\begin{centering}
    \begin{tabular}{ | l | l | l |}
    \hline
    {\bf Data Set} & {\bf \# Queries} & {\bf Similarity Measure} \\ \hline
    \texttt{test.m} & 79 & Disco BNC \\ \hline
    \texttt{bill.m} & 131 & Disco Wikipedia \\ \hline
    \texttt{rami.m} & 305 & Disco Wikipedia \\ \hline
    \texttt{nikhil.m} & 600 & Disco Wikipedia \\ \hline
    \end{tabular}
\end{centering}
\bigskip

As you can see, our data sets, intentionally, vary greatly in size
(number of queries). This allows us to test the effectiveness of our
classification strategy in both small and large window sizes. Data in
the larger sets span multiple sessions over multiple days. The
\texttt{test.m} and \texttt{bill.m} data sets are taken from a user's
search engine use over a single day.

To validate our classifier, we first compute a similarity matrix
using the Disco Wikipedia distance measure. From this matrix, we
performed clustering using PAM, as described in Section
\ref{sec:clustering}. Our classifier then identified the largest
cluster or set of clusters, assigning the label `U' to these
queries. We finally compare the assigned label against the real value
to calculate the precision and recall over our validation data set.

\section{Results}
\label{sec:results}

Our results are very promising. We are able to successfully identify a
subset of the user generated queries with a high precision for small
window sizes. It is important to note that we make no attempt to
classify all user queries. Thus, our recall is very low, even for
small window sizes.

\bigskip
\begin{centering}
    \begin{tabular}{ | l | l | l |}
    \hline
    {\bf Data Set} & {\bf Precision} & {\bf Recall} \\ \hline
    \texttt{test.m} & 1.0 & 0.147 \\ \hline
    \texttt{bill.m} & 1.0 & 0.25 \\ \hline
    \texttt{rami.m} & 0.867 & 0.188 \\ \hline
    \texttt{nikhil.m} & 0.125 & 0.067 \\ \hline
    \end{tabular}
\end{centering}
\bigskip

In addition to classification, our tool outputs what we term a cluster
map. It is a way of organizing and visualizing the similarity matrix
by cluster assignment. Clusters are organized by size, starting with
the largest at the top left corner, to the smallest in the bottom
right corner. Any position, $(i,j)$, represents the similarity between
query $i$ and query $j$. The darker the map is at that index, the
farther apart the two queries are. Additionally, the true class label
is displayed along the diagonal, replacing the trivial similarity of 0
between a query, $i$, and itself.

\begin{figure*}[h]
    \centering
    \includegraphics[width=\linewidth]{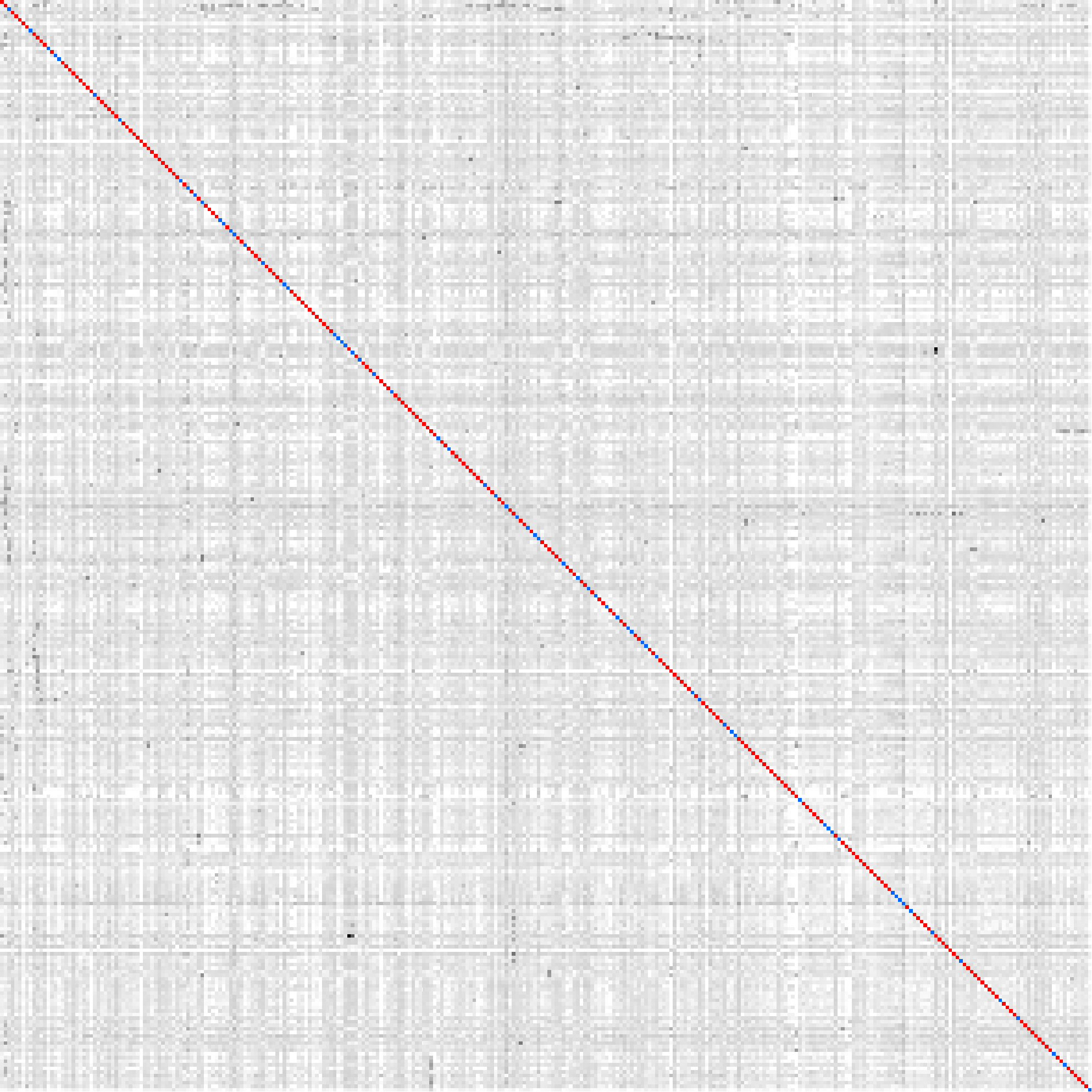}
    \caption{A pictorial representation of the \texttt{rami.mat}
      clustering using GND. Position $(i,j)$ represents the similarity
      between query $i$ and query $j$ - the lighter the shade, the
      more similar the queries are. Each query's class is displayed
      along the diagonal. If $(i,i)$ is red, then query $i$ was
      generated by TrackMeNot. If $(i,i)$ is blue, then the query is
      genuine. We see that the similarity matrix itself is
      homogeneous; most queries are equally semantically
      unrelated. Only one cluster was found, spanning the entire data
      set. We conclude that GND is not an appropriate similarity
      measure when applied to entire query strings.}
    \label{fig:rami.ngd}
  \end{figure*}

We also performed cluster analysis of the data using Google Normalized
Distance as our measure of semantic similarity. The \texttt{rami.m}
analysis is shown in Figure \ref{fig:rami.ngd}. We find that this
measure is not appropriate for calculating similarity at the scale of
entire queries. Like Disco, GND is capable of computing distances
between keyword pairs only. When inputing entire queries, the similarity
matrix is homogeneous; most queries are comparably close/far from most
other queries, and no clusters are found.

  \begin{figure*}[h]
    \centering
    \includegraphics[width=\linewidth]{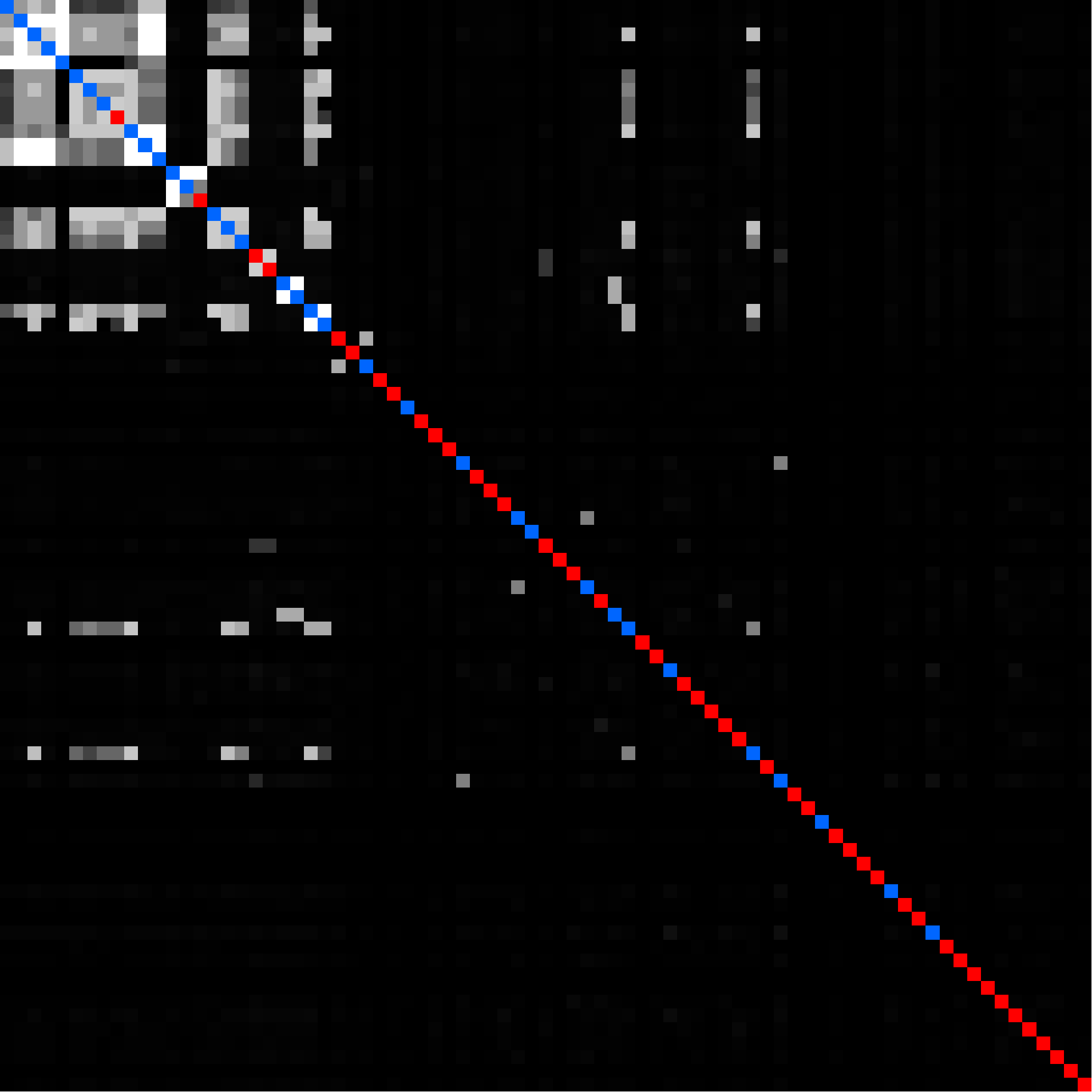}
    \caption{A pictorial representation of the \texttt{test.m}
      clustering. Position $(i,j)$ represents the similarity between
      query $i$ and query $j$ - the lighter the shade, the more
      similar the queries are. Each query's class is displayed along
      the diagonal. If $(i,i)$ is red, then query $i$ was generated by
      TrackMeNot. If $(i,i)$ is blue, then the query is genuine. We
      see that many queries are semantically unrelated. The largest
      cluster, seen in the top left of the graph, has 5 user queries and 0
      TrackMeNot queries.}
    \label{fig:disco.test}
  \end{figure*}

  \begin{figure*}[h]
    \centering
    \includegraphics[width=\linewidth]{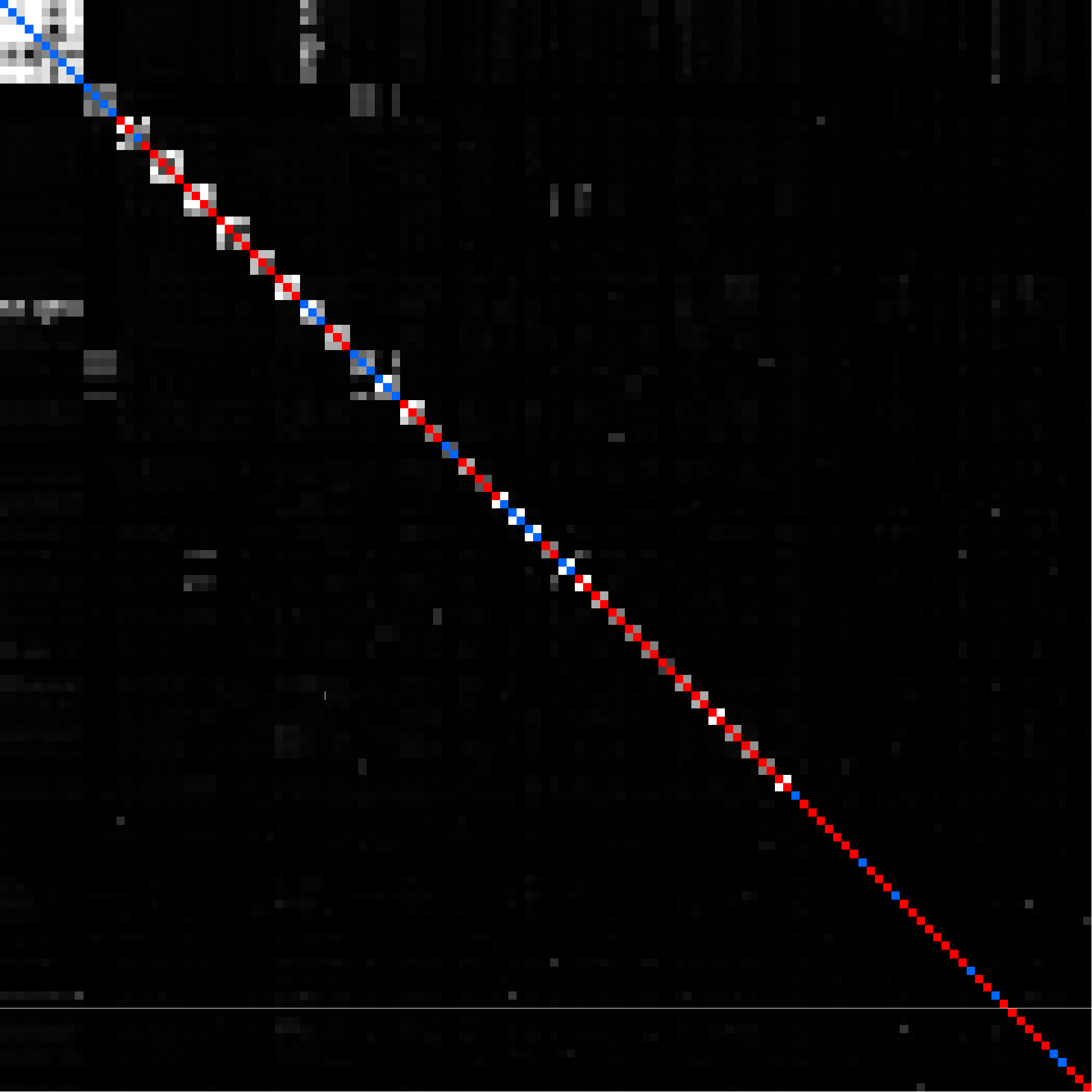}
    \caption{A pictorial representation of the \texttt{bill.m}
      clustering. Position $(i,j)$ represents the similarity between
      query $i$ and query $j$ - the lighter the shade, the more
      similar the queries are. Each query's class is displayed along
      the diagonal. If $(i,i)$ is red, then query $i$ was generated by
      TrackMeNot. If $(i,i)$ is blue, then the query is genuine. We
      see that many queries are semantically unrelated. The largest
      cluster, seen in the top left of the graph, has 10 user queries
      and 0 TrackMeNot queries.}
    \label{fig:disco.bill}
  \end{figure*}

  \begin{figure*}[h]
    \centering
    \includegraphics[width=\linewidth]{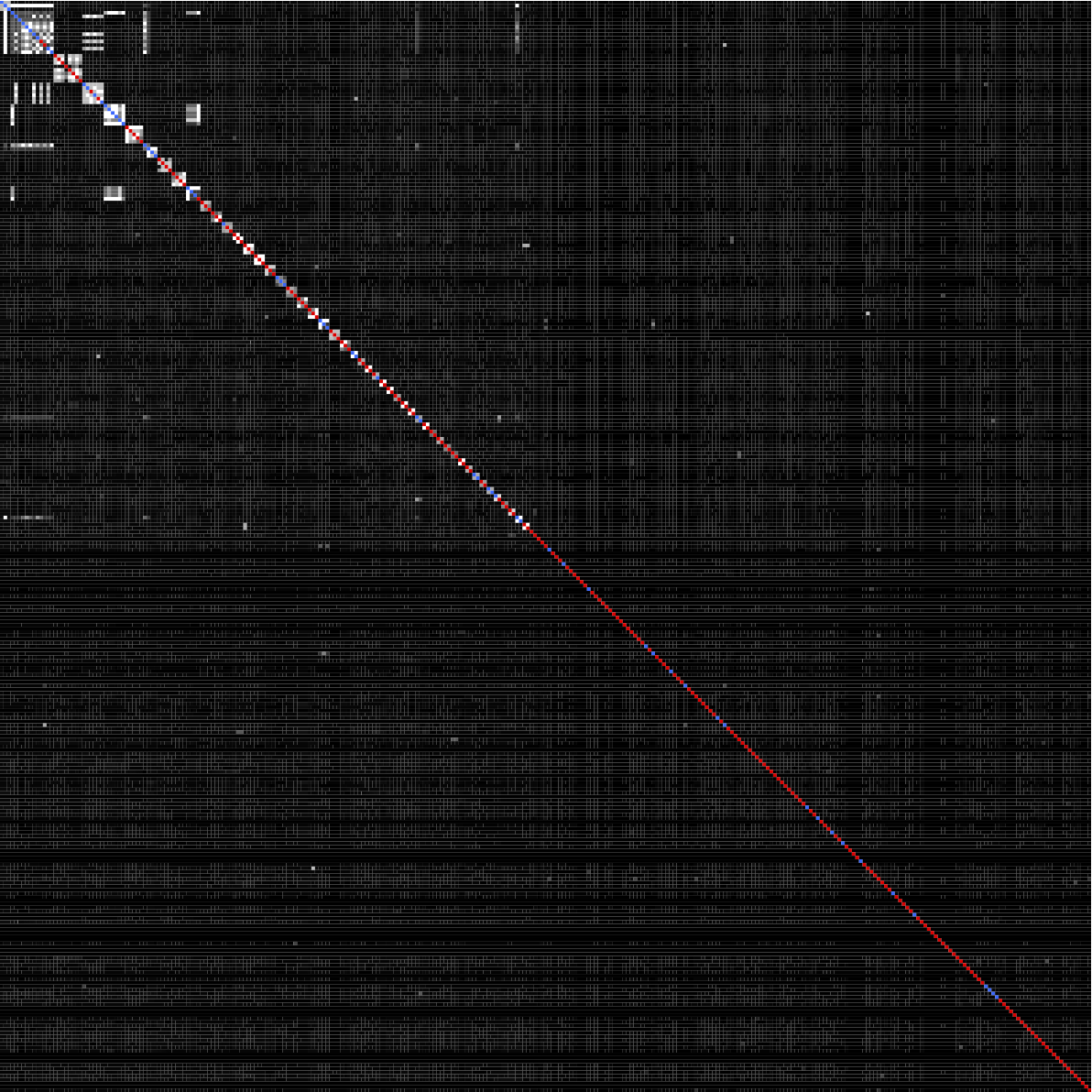}
    \caption{A pictorial representation of the \texttt{rami.m}
      clustering. Position $(i,j)$ represents the similarity between
      query $i$ and query $j$ - the lighter the shade, the more
      similar the queries are. Each query's class is displayed along the
      diagonal. If $(i,i)$ is red, then query $i$ was generated by
      TrackMeNot. If $(i,i)$ is blue, then the query is genuine. We
      see that many queries are semantically unrelated. The largest
      cluster, seen in the top left of the graph, has 13 user queries and 2 TrackMeNot queries.}
    \label{fig:disco.rami}
  \end{figure*}

  \begin{figure*}[h]
    \centering
    \includegraphics[width=\linewidth]{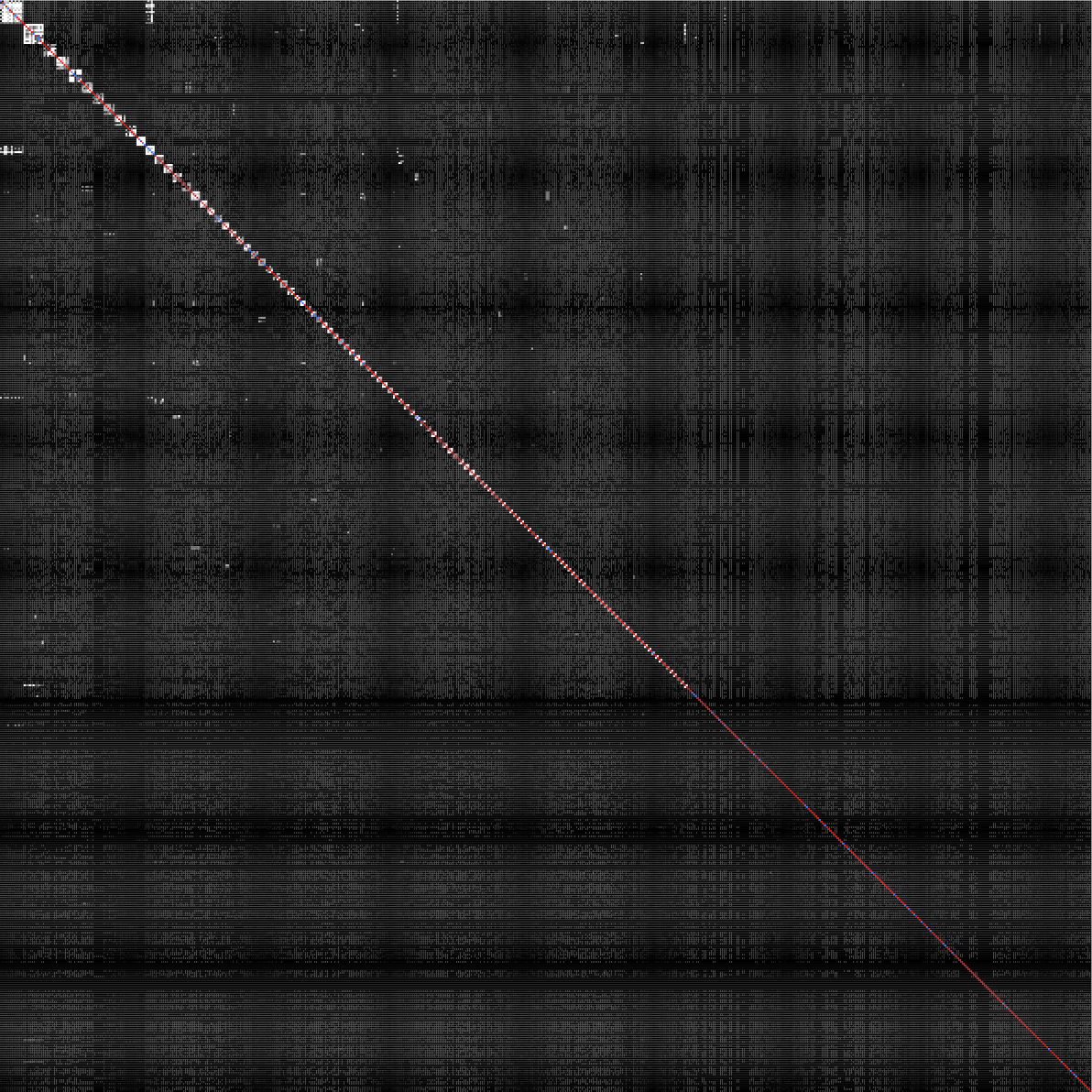}
    \caption{A pictorial representation of the \texttt{nikhil.m}
      clustering. Position $(i,j)$ represents the similarity between
      query $i$ and query $j$ - the lighter the shade, the more
      similar the queries are. Each query's class is displayed along
      the diagonal. If $(i,i)$ is red, then query $i$ was generated by
      TrackMeNot. If $(i,i)$ is blue, then the query is genuine. We
      see that many queries are semantically unrelated. The largest
      cluster, seen in the top left of the graph, has 4 user queries
      and 9 TrackMeNot queries.}
    \label{fig:disco.nikhil}
  \end{figure*}

\section{Conclusions}
\label{sec:conc}

For small windows, our classifier performs well. We achieve perfect
precision in the \texttt{test.m} and \texttt{bill.m} data sets. But as
our window size grows, we observe declining precision. 

Intuitively this decline makes sense.  TrackMeNot does seed its
dynamic query list with multiple entries from an individual source, so
we expect to find multiple groups of related searches within its query
list at any given time. The reason our classifier performs well for
small windows is that TrackMeNot chooses its queries at random. As the
window size increases, so does the coverage of TrackMeNot's query
list. We do not see poor performance for large windows as a cause for
concern. Rather, it confirms the reasoning behind our strategy.

We also find that the Google Normalized Distance did not serve as a good
distance measure when applied to full queries. If we instead used it
per word in each query, and aggregated the distance for each word in
each query like in DISCO, we could get a better measure.

\section{Future work}
\label{sec:future}
As seen in our results, TrackMeNot does succeed in protecting the
user's privacy over a long period of time by forming clusters even in
the set of TrackMeNot queries. However, TrackMeNot does not protect
against search engines tracking time of use. For example, as also
shown in our results, if a clustering of Google Search queries is
performed over smaller time-frames, it is effective in filtering
the TrackMeNot queries from those of the user because in smaller
time-frames, TrackMeNot queries do not form big enough
clusters. Hence, we feel that by incorporating more semantic
relatedness in consecutive TrackMeNot queries, then even in smaller time-frames, the
queries generated by TrackMeNot will form clusters that will make it
difficult to filter them out using a clustering analysis such as ours.

Also, by operating as a Firefox plugin, TrackMeNot operates only when
the user has Firefox running. This prevents TrackMeNot from firing off
queries when the user is not actively browsing the Internet. We are
convinced that if TrackMeNot can run as a standalone application
running continuously in the background, this would not only provide
the same semantic search noise as the current TrackMeNot
implementation, but also serve to protect the temporal search habits
of users.

\bibliographystyle{abbrv}
\bibliography{final_report}
\end{multicols}
\newpage

\end{document}